\documentclass[%
 reprint,
superscriptaddress,
 amsmath,amssymb,
 aps,
]{revtex4-1}

\usepackage{CJKutf8}

\usepackage{graphicx}
\usepackage{stmaryrd}
\usepackage{amssymb,amsmath,amsfonts,latexsym,mathrsfs,amsthm}
\usepackage{hyperref}\hypersetup{colorlinks=true,linkcolor=blue,citecolor=blue}

\newcommand{\cref}[0]{C_{\rm{ref}}}
\newcommand{\vb}[0]{V_{\rm{b}}}
\newcommand{\qtot}[0]{Q_{\rm{tot}}}
\newcommand{\subl}[1]{{#1}_{\rm L}}
\newcommand{\subh}[1]{{#1}_{\rm H}}

\newcommand{\jump}[1]{\left\llbracket{#1}\right\rrbracket}

\begin{document}


\title{Battery Detached Energy Conversion by Pyroelectric Effect}
\begin{CJK*}{UTF8}{gbsn}
\author{Chenbo Zhang (张晨波)}
\affiliation{Department of Mechanical and Aerospace Engineering, Hong Kong University of Science and Technology, Clear Water Bay, Hong Kong}
\author{Yintao Song (宋寅韬)}
\affiliation{Independent Researcher, 1170 Foster City Blvd, Foster City, CA 94404, USA}
\author{Maike Wegner}
\author{Eckhard Quandt}
\affiliation{University of Kiel, Faculty of Engineering, 24143 Kiel, Germany}
\author{Xian Chen (陈弦)}
 \email{xianchen@ust.hk}
\affiliation{Department of Mechanical and Aerospace Engineering, Hong Kong University of Science and Technology, Clear Water Bay, Hong Kong}

\date{\today}

\begin{abstract}
We propose a pyroelectric energy conversion device that converts heat directly to electricity. 
In contrast to conventional pyroelectric energy conversion designs, this energy harvesting system is detached from any external power sources, operating only under periodically varying temperature. 
Such detachment unambiguously attributes the converted electricity to heat that drives the change of polarization in the pyroelectric material, not to the electric field alternation caused by the external battery. Using pure and Zr doped BaTiO$_3$, we demonstrate the electricity generation in consecutive temperature cycles. 
We further develop a thermodynamic model for the energy conversion system. Our model suggests that the work output is rate dependent: the work output per cycle is linearly dependent on the heat/cooling frequency below the predicted threshold. 
The linearity is confirmed by experiments, and the threshold frequency is derived by theory. 
Finally we propose a figure of merit that separates the materials intrinsic properties from the system design parameters. 
The figure of merit guides the future material development and device improvement.
Our work clears out confusions and reforms the foundation for pyroelectric materials' resurgence as a competitor for green electricity.
\end{abstract}


\maketitle
\end{CJK*}


\section{Introduction}

According to the 2008 annual report of U.S. Department of Energy, one third to a half of the industrial energy input is wasted as heat of temperature around the boiling point of water \cite{johnson2008waste}.
Recycling such low grade waste heat and integrating it for power generation have a great impact on both environment and economy.
By far, the thermoelectrics, as the most successful direct power harvesting devices from heat, have dominated the applications in this area \cite{zhang2015,zhao2014}.
However, their efficiency and work output density are hindered by the small $ZT$ (the thermoelectric figure of merit) values at the low temperature regime \cite{tan2016non,he2014high}.
A recently emerging branch of methods for direct energy conversion from heat is to manipulate the caloric effect with one of the multiferroic properties, e.g the pyroelectric effect of ferroelectric materials \cite{Pandya2018} and the magnetocaloric effect of ferromagnetic materials \cite{Srivastava2011, Song2013}.
Generally, the magnetoelectric properties of a material are sensitive to change in lattice parameters due to, for example, heating and cooling. 
This mechanism underlies the temperature-dependent ferroic properties.
When these temperature dependent ferroic properties are coupled with a structural phase transformation, the caloric effects are dramatically amplified \cite{Wada2001, Moya2013, Moya2014, Zhao2017}, which makes the applications of energy harvesting admissible and practical. 

For ferroelectric materials, the large pyroelectric effect is commonly observed during a structural transformation between ferroelectric phase and paraelectric phase.
In most of the cases, such a structural transformation is martensitic, that is a diffusionless solid-solid reversible phase transformation.
As a result, within a narrow temperature range, the polarization of the material exhibits an abrupt jump.
State-of-the-art technologies of fabricating sophisticated devices have demonstrated promising pyroelectric properties that can be leveraged by this kind of energy harvesting. \cite{Lee2013, Pandya2018}
For examples, within 10 K temperature range, the polarization jump $\jump{P} = 6$ $\mu$C/cm$^2$ in single crystal BaTiO$_3$ \cite{Moya2013} and $5.5$ $\mu$C/ cm$^2$ in ferroelectric PMN-$0.32$PT thin films. \cite{Pandya2018}
These materials are good candidates for this kind of energy harvesting device.

The first pyroelectric energy device was designed as a capacitor. The working mechanism is based on 
the different kinds of quasistatic thermodynamic cycles proposed in early 80s \cite{Olsen1983}.
Among them, the Ericsson cycle, also known as the Olsen cycle in pyroelectric literature, shows the maximum energy output converted directly from heat \cite{Olsen1985}.
In this device, the dielectric layer of the capacitor is made of a phase-transforming ferroelectric material that generates electric current due to the change of polarization when being heated across the phase transformation.
Using the Ericsson cycle, the relationships between the pyroelectric properties and the energy harvesting performance were widely studied in various ferroic materials systems. \cite{Olsen1983, Olsen1985, Sebald2006, Lee2013, Pandya2018}

An Ericsson cycle consists of two isothermal processes at the temperatures below and above the transition temperature and two isobaric processes under different applied electric fields.
Energy conversion by pyroelectric effect utilizing the Ericsson cycle has been demonstrated in many material systems using similar designs. \cite{Olsen1985, Lee2013, Pandya2018}
The common idea of these designs is to connect the pyroelectric capacitor to a load resistor and a battery in series. The load resistor serves as an electricity consumer, across which the changes in voltage/current is considered as the harvested energy from the device.
The role of the battery is to provide the device with a bias voltage which controls the change in applied field during the cycles.
During the isothermal process at the low temperature, the device is in ferroelectric phase spontaneously polarized by the applied bias voltage.
Holding the electric field, the device is heated up through the transformation temperature, simultaneously its capacitance jumps due to the phase transition.
As a result, the capacitor releases charges passing through the resistor and generates electricity.
The measure of the voltage on the load resistor connected to the capacitor is commonly considered as the energy harvested during the heating process. \cite{Lee2013,Pandya2018}
At the end of the phase transformation, either the system goes through a symmetric cooling half cycle, or the resistor is short-circuited by a diode and then the system restores to its original state through an isothermal discharging and an isobaric cooling.\cite{Olsen1985}
A schematic diagram is sketched in FIG. \ref{fig:schematics}a, with the optional diode omitted.

In above conventional designs for pyroelectric energy conversion,  \cite{Olsen1985, Sebald2006, Lee2013, Pandya2018} the electrostatic energy converted from heat is often confused with the electricity collected on the resistor.
These two quantities are not necessarily equal.
In fact, any variational electrical field can generate a current in a resistor-capacitor circuit.
We demonstrate this by replacing the pyroelectric capacitor in conventional design (FIG. \ref{fig:schematics}a) with a regular one, and simply toggle the power source voltage between 10 and 30 V. 
Without surprise, we observe electricity signal, indicated by voltage changes, on the resistor, as shown in FIG. \ref{fig:schematics}b. 
Therefore when analyzing such energy conversion system, one needs to be very careful how much electricity collected on the resistor is attributed to the pyroelectric effect rather than the variation of the external power source.
In particular, electricity signals observed in the isothermal processes should not be considered as the power generated from heat. However,
the influence of such signals is unavoidable in designs that are chargeable by external power source during operation.
Motivated by this, in this paper we propose a battery detached device: no external power source is involved while the device is converting energy from heat to electricity.

Another factor often overlooked in thermodynamic analysis of pyroelectric energy conversion is the rate dependency of work output. 
Modern theoretical framework of thermodynamics is built on top of Carnot's work \cite{carnot1824reflexions} for heat engine.
If we analogue a common pyroelectric energy conversion device to a traditional piston-cylinder heat engine, the energy collected on the load resistor is like the heat generated due to the friction between the piston and the cylinder wall. This type of energy is always considered as dissipation.
Therefore the pyroelectric energy conversion is rate dependent, which is however ignored in most classical thermodynamic models.
In this paper, we discuss the rate dependency of work output in our battery detached design.
To consider the performance, the energy conversion designs by pyroelectric effect, including our battery detached one, are subjected to many coupled parameters,
including material constants, operation conditions, and circuit design parameters.
By nondimensionalization of the governing equation, we identify a threshold cycling frequency under which the work density varies linearly with the heating/cooling rate. In the end, 
we propose 
a simple figure of merit underlying the performance of a pyroelectric energy conversion system using our design.

\section{Battery detached pyroelectric energy conversion from heat to electricity}
\begin{figure}[ht]
\centering
\includegraphics[width=3in]{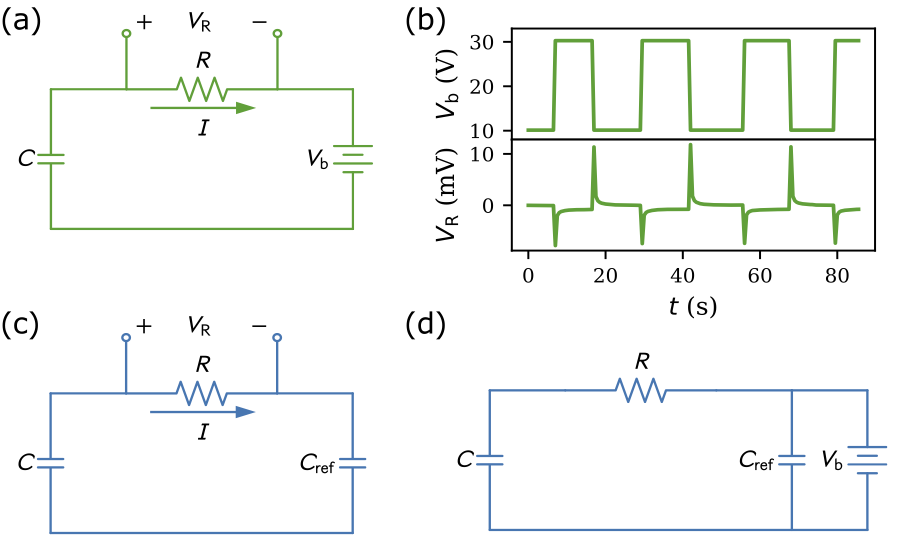}
\caption{Schematics of conventional and new battery detached pyroelectric energy conversion circuits. (a) is the schematic diagram of the conventional design based on Olsen's design \cite{Olsen1985}. (b) demonstrates that in the conventional design even using a regular capacitor, simply by varying the battery voltage, one can collect electricity signal on the resistor, which is clearly not converted from any heat source. (c) is the schematic diagram of our battery detached design. (d) is the circuit of the one time initial charging of both capacitors at room temperature before removing the battery and starting the cyclic heating and cooling.}\label{fig:schematics}
\end{figure}

In our design, we replace the battery in conventional design by a reference capacitor ($C_\text{ref}$) as shown in FIG. \ref{fig:schematics}c. 
Initially when the temperature of pyro-capacitor is at $\subl T < \theta$ where $\theta$ is the phase transformation temperature, the pyro-capacitor and the reference capacitor are both fully charged by the external power source of voltage $\vb$ so that no charges flow in the circuit.
This is counted as the initial state of the device shown in FIG. \ref{fig:schematics}d.
Afterwards, the external power source is detached and the total injected charges in the circuit is assumed to be preserved.
\footnote[4]{In practice there will be some charge losses due to leakage, which will not be discussed in this paper.}
A resistor with resistance $R$ is connected in series to the pyroelectric capacitor to detect the current generated during the cyclic heating/cooling processes.
The produced voltage across the resistor is characterized as $V_{\rm R}$.
This value will be the index of our energy conversion circuit as it signals the energy converted purely from the phase transformation.
In order to make the system alternate between $\subl T$ and $\subh T$ cyclically, 
we maintain two heat reservoirs provide stable temperatures of $\subl T$ and $\subh T$ respectively.
As a proof of concept, the phase transformation is driven by manually moving the pyroelectric capacitor between the two heat reservoirs. 

\begin{figure}[ht]
\centering
\includegraphics[width=3.15in]{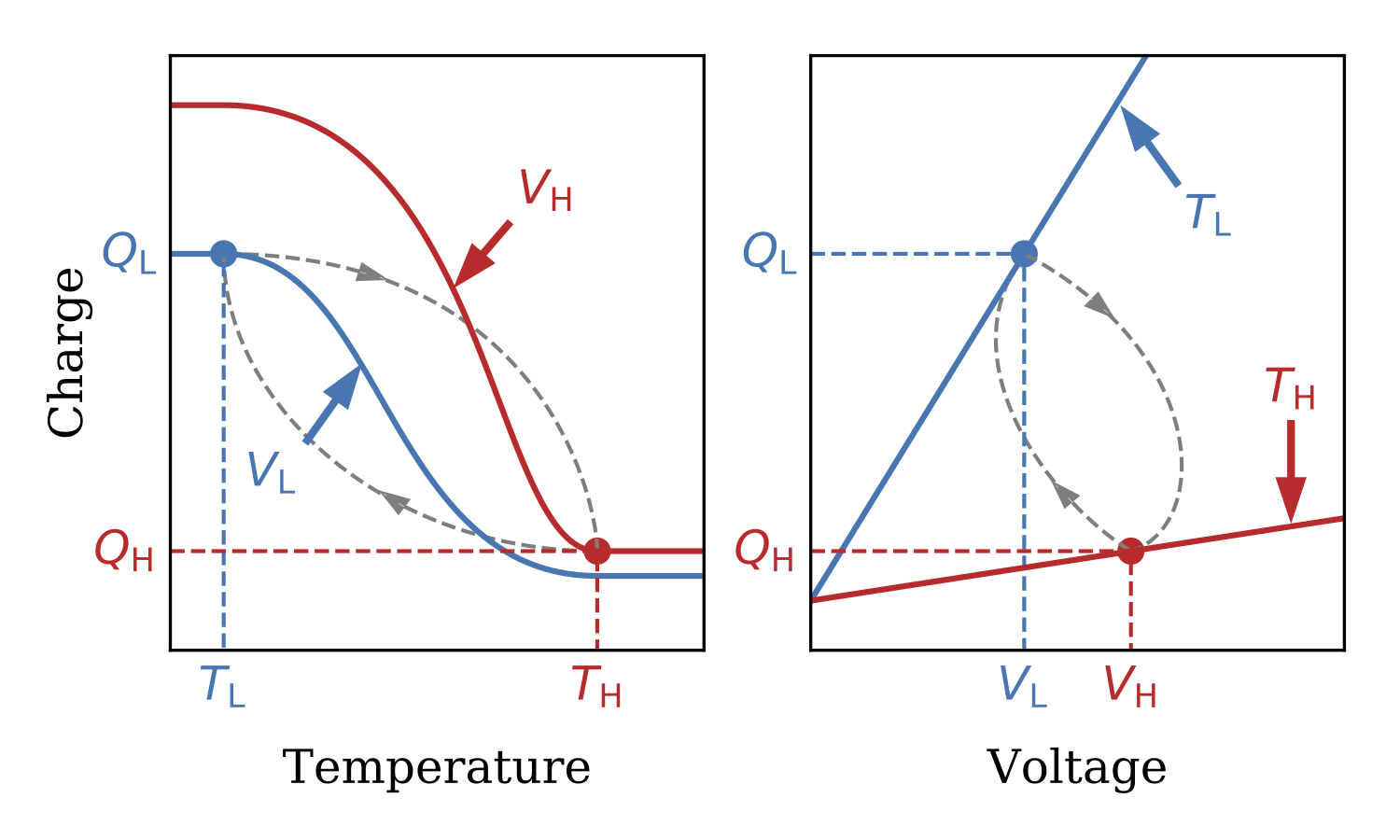}
\caption{Sketch of thermodynamic states of a pyroelectric capacitor}\label{fig:qt}
\end{figure}

Let the pyroelectric capacitor be the thermodynamic system. The free charge on its surface is a function of voltage $V$ and temperature $T$, i.e. $Q(V, T)$, sketched in FIG. \ref{fig:qt}. This is regarded as the key physical property responsible for energy conversion. By the charge conservation of the energy conversion circuit in FIG. \ref{fig:schematics}c, when the system jumps between the ferro- and para-electric phases, the capacitance of the system alters accordingly, which drives the movement of free charges between the two capacitors. Note that since $Q(\subl V, \subl T) > Q(\subh V, \subh T)$, the system will be discharged/charged spontaneously during heating/cooling processes without any external bias voltage.  
Unlike any of the quasi-static thermodynamic cycles assumed by Olsen \cite{Olsen1985}, both voltage and temperature of the system continuously change at the same time along the heating/cooling paths between the two states $(\subl Q, \subl V)$ and $(\subh Q, \subh V)$ in $Q-V$ plane, where $Q_{\rm L, H}$ and $V_{\rm L, H}$ are the charges and the corresponding voltages of the pyroelectric capacitor at the low and high temperature phases. 

\section{Thermodynamic model}
\subsection{Governing equation of energy conversion}
In order to have a rational understanding of the relationships among the material intrinsic properties, design parameters and control parameters of our battery-detached energy conversion device, we use the electric charge and voltage as the energy conjugate to study the thermodynamics of the system. 
The dynamics of the charge held by the pyroelectric capacitor is modeled by the time dependent voltage $V(t)$ and temperature $T(t)$. That is, at $t \geqslant 0$, 
$Q(t) = Q(V(t), T(t))$ with the initial voltage $V(0) = \vb$ and temperature $T(0) = \subl T$. The function $Q(V, T)$ is a state function that can be determined by the state variables $V$ and $T$ despite of the time history. 

The initial voltage $\vb$ of the system and the capacitance of the reference capacitor $\cref$ determine the total charges in the circuit,  
\begin{equation}\label{eq:qtot}
\qtot = \cref V_{\rm b} + Q(V_{\rm b}, \subl T).
\end{equation}
When the system reaches the steady state in ferroelectric phase at $T = \subl T < \theta$ after being detached from the external power source,
the corresponding voltage and charge are the same as the initial state   
\begin{equation}
\subl V = V_{\rm b}, \quad \subl Q = Q(V_{\rm b}, \subl T).    
\end{equation}
Upon heating, the system transforms from ferroelectric phase to paraelectric phase.
When the phase transformation finishes, the system reaches the other steady state at $T = \subh T$.
By charge conservation, the corresponding voltage and charge are then
\begin{equation}
\subh V = \frac{\qtot - \subh Q}{\cref}, \quad \subh Q = Q(\subh V, \subh T).
\end{equation}
In the temperature range of $\subl T < T < \subh T$, the charges flow in the circuit due to the change of the capacitance of the system.
By Kirchhoff's law of voltage,
\begin{equation}\label{eq:law-of-volt}
V = -\dot{Q}(V, T) R + \frac{\qtot - Q(V, T)}{\cref}.
\end{equation}
In equation \eqref{eq:law-of-volt}, the parameter $R$ denotes the resistance of the resistor across which the output voltage $V_{\rm R} = -\dot{Q}(V, T) R$ is used to measure the energy conversion over the cyclic thermal loading. 

$T(t)$ is mainly governed by heat transfer between the system and heat reservoirs.
In our experiment, we can tune the $T(t)$ by manipulating the temperatures of the two reservoirs to achieve different heating/cooling rates. 
In theoretical analysis, we treat them as external stimuli.
Then \eqref{eq:law-of-volt} becomes a first order nonlinear ordinary differential equation (ODE) for $V(t)$:
\begin{equation}\label{eq:ode}
    R\cref\dot{Q} + \cref V + Q - \qtot = 0.
\end{equation}

\subsection{Change of variables}
Using the approximation of the planar capacitor with area $A$ and thickness $d$ satisfying $d \ll \sqrt{A}$, the free charges accumulated at the surface of the pyroelectric capacitor is estimated as 
\begin{equation}\label{eq:q}
    Q(V, T) = \epsilon_0 E A + P(E, T) A,
\end{equation}
where $\epsilon_0$ is the vacuum permittivity and the electric field $E$ in the capacitor is calculated as
\begin{equation}
E = \frac{V}{d}.    
\end{equation}

The constitutive function $P(E, T)$ is the polarization of the pyroelectric material under the applied electric field $E$ at the temperature $T$. This is an intrinsic material property that can be characterized by measuring the $P-E$ curves at different temperatures \cite{Moya2013}. To link the thermodynamic behavior of the system to the material property $P$, we change the energy conjugate $(Q, V)$ to $(P, E)$, and the governing equation \eqref{eq:ode} becomes a first order nonlinear ODE for the electric field in the pyroelectric material $E(t)$.
The total charge \eqref{eq:qtot} can be rewritten as
\begin{equation}\label{eq:qtot-2}
    \qtot = \cref \vb + \frac{\epsilon_0\vb A}{d} + \subl P A.
\end{equation}
The parameter $\subl P$ represents the polarization of the system at the low temperature phase, i.e.
\begin{equation}
\subl P = P\left(\frac{V_{\rm b}}{d}, \subl T\right).    
\end{equation}
Note that the value of $\subl P$ can be experimentally characterized by the $P$-$E$ curve at temperature $\subl T$.

\subsection{Nondimensionalization of the governing equation}\label{sec:nondim}
Eq. \eqref{eq:ode} is a nonlinear ODE.
It can be numerically integrated.
In order to uncover the intrinsic relationships among coupled parameters, we convert the governing equation into a dimensionless form.
We use bar $\bar \cdot$ to denote dimensionless variables.

Choose $V_{\rm b}$ as the characteristic voltage, then
\begin{equation}
V = V_{\rm b} \bar V, \quad E = \frac{V_{\rm b}}{d} \bar E.    
\end{equation}
We scale polarization and temperature in the following way:
\begin{equation}
P = \subl P \bar P, \quad
T = \frac{\subh T - \subl T}{2}\bar T + \theta\left(\frac{\vb}{d}\right).    
\end{equation}
Here, the function $\theta(E)$ denotes the transformation temperature under the electric field $E$. A closed form will be proposed in the discussion of the constitutive assumption in Section \ref{sec:cons}.  In principle, the phase transformation temperature of a material varies with the application of external load, which is underlain by the Clausius-Clapeyron relation. 

Define dimensionless charge as
\begin{equation}
Q = \subl P A \bar Q.    
\end{equation}
Combining with \eqref{eq:q}, we get
\begin{equation}
\bar Q = \frac{\epsilon_0 \vb}{\subl P d}\bar E + \bar P.
\end{equation}
Finally, time is scaled by 
\begin{equation}
t = \tau \bar t,    
\end{equation}
in which $\tau$ is the duration of heating half cycle. 
We consider this time scale to be a given constant in our model, because it is mostly governed by heat transfer and the thermal property of the material.
As a rough estimate, if the intake heat flow is $q$ and the latent heat of phase transformation is $\ell$, the duration for completely transforming the material from low temperature to high temperature phase is $\ell / q$.

Substituting equation \eqref{eq:qtot-2} into \eqref{eq:ode} and nondimensionalizing it, we have
\begin{equation}\label{eq:ode-bar}
\frac{R\cref}{\tau}\dot{\bar Q} + \bar Q + \frac{\vb\cref}{\subl P A} (\bar E - 1)
- \left(\frac{\vb\epsilon_0}{\subl Pd} + 1\right)= 0.
\end{equation}
Equation \eqref{eq:ode-bar} does not assume any particular form of constitutive response $\bar P(\bar E, \bar T)$, or equivalently $P(E, T)$.

Solve ODE \eqref{eq:ode-bar} for $\bar E(\bar t)$, then we can reconstruct the dimensional current and the voltage on resistor as
\begin{equation}
I(t) = -\dot Q(t) = - \frac{\subl P A}{\tau}\dot{\bar Q}{\rm,}\quad
V_{\rm R}(t) = I(t)R.    
\end{equation}
We assume the duration of the cooling half cycle is also $\tau$. The total energy harvested during a full cycle started at $t_0$ is 
\begin{equation}\label{eq:work}
{\cal W} = \int_{t_0}^{t_0 + 2\tau} I^2 R {\rm d}t
= \frac{\subl P^2 A^2 R}{\tau} \int_{\bar t_0}^{\bar t_0 + 2} \dot{\bar Q}^2{\rm d}\bar t.
\end{equation}

\section{Experiment and simulation}

\subsection{Energy conversion demonstration}
We prepared three thin plate specimens: (i) pure BaTiO$_3$ (BTO) with thickness of $0.77$ mm and area of $62.35$ mm$^2$, (ii) BaZr$_{0.006}$Ti$_{0.994}$O$_3$ (Zr$_{0.006}$) with thickness of $0.58$ mm and area of $69.54$ mm$^2$, and (iii) BaZr$_{0.01}$Ti$_{0.99}$O$_3$ (Zr$_{0.01}$) with thickness of $0.57$ mm and area of $33.12$ mm$^2$.
The bulk ceramics were prepared by a conventional solid state reaction route. 
The appropriate amounts of BaCO$_3$, TiO$_2$ and ZrO$_2$ according to the target stoichiometry were weighed and mixed with a planetary mill in n-Hexane with Zirconia balls. 
The calcination of the powders was performed twice at 1350 $^\circ$C. 
After each calcination step the samples were ball milled. 
The obtained powders were mixed with a binding agent and pressed into discs with a diameter of 20mm. 
For sintering the pressed discs were surrounded by BaTiO$_3$-granulate to get a homogeneous temperature distribution within the specimen. 
The binder was burnt out at 500 $^\circ$C for 1 hour, afterwards the temperature was set to 1500 $^\circ$C for 3 hours for the sintering process.

\begin{figure}[ht]
\centering
\includegraphics[width=3in]{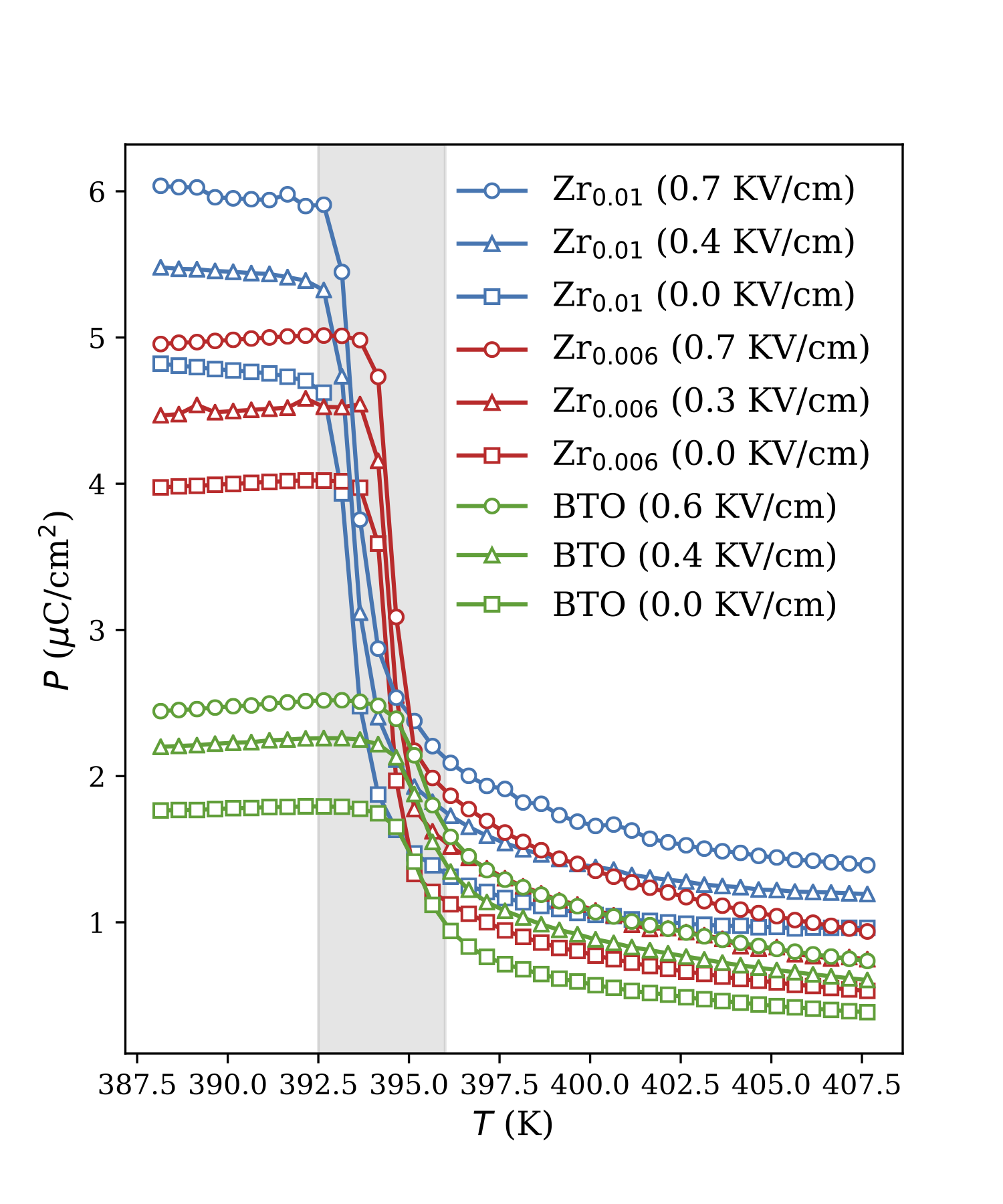}
\caption{Pyroelectric property of materials. Shaded area indicates the temperature range for phase transformation. \label{fig:pyroelectric}}
\end{figure}

The pyroelectric property in all three specimens were characterized by an aixACCT TF Analyzer 2000E. 
At a given temperature, the electric field varies between $\pm10$ kV/cm at the frequency of 50 Hz.
The value of $P(E, T)$ is evaluated at the upper branch of the hysteresis loop. \cite{Moya2013}
As shown in FIG. \ref{fig:pyroelectric}, the specimen Zr$_{0.01}$ has the largest jump in polarization across the phase transformation. The Zr doped BaTiO$_3$ specimens show bigger $d P/d T$ than the pure BaTiO$_3$.
The ferro- to para-electric phase transformation in all specimens occurs in the temperature range between 392.5 K and 396.0 K, marked as the shaded area in the picture.

\begin{figure}[ht]
\centering
\includegraphics[width=3in]{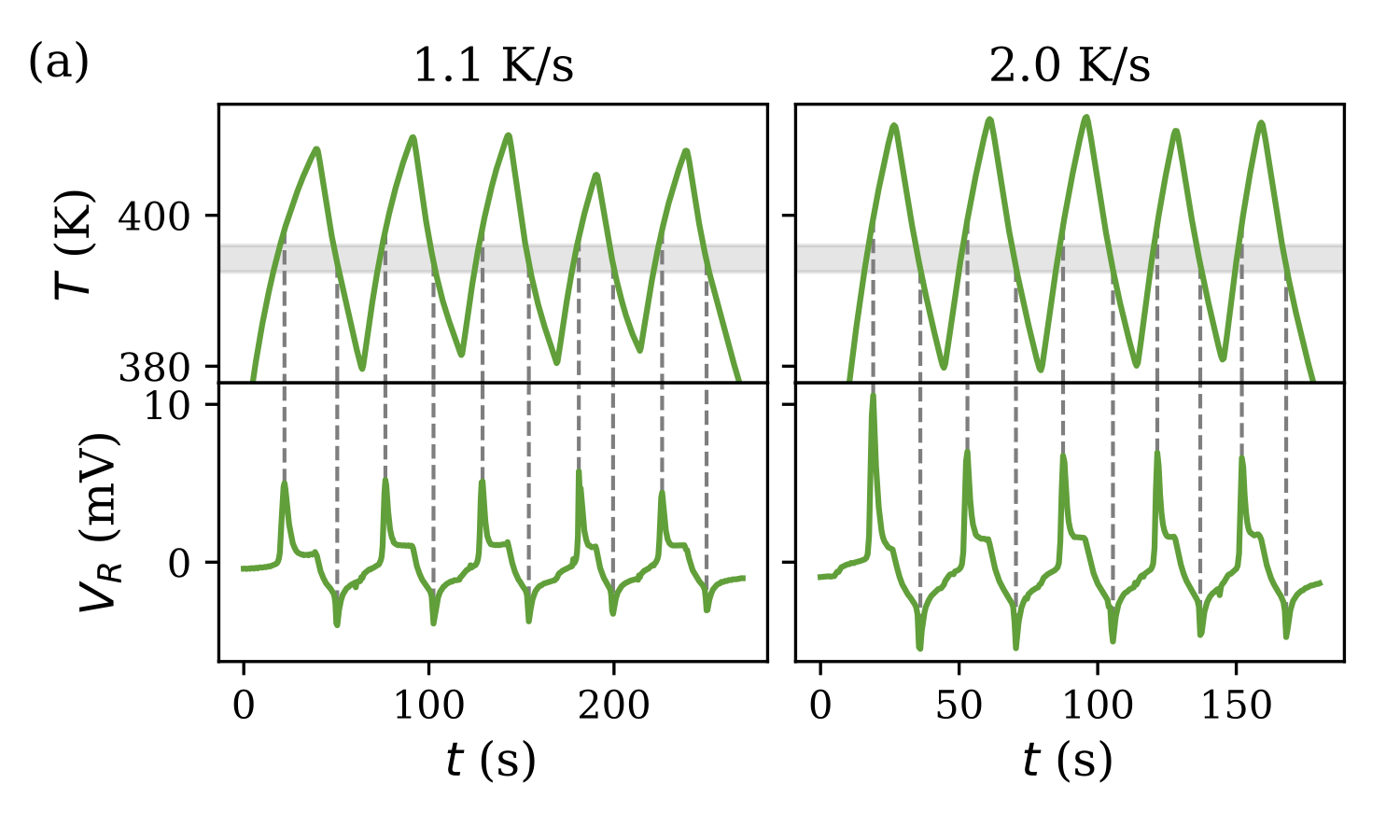}\\
\includegraphics[width=3in]{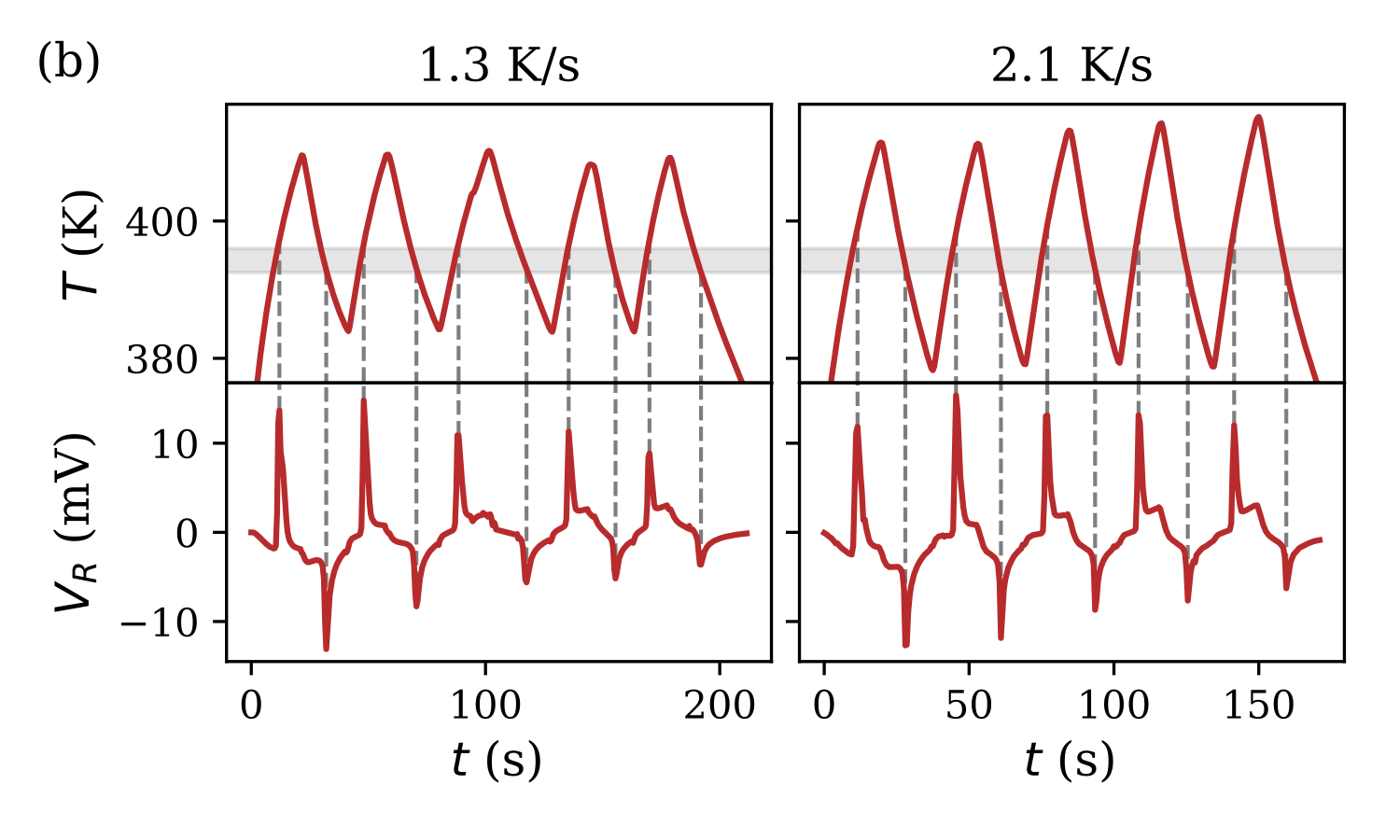}\\
\includegraphics[width=3in]{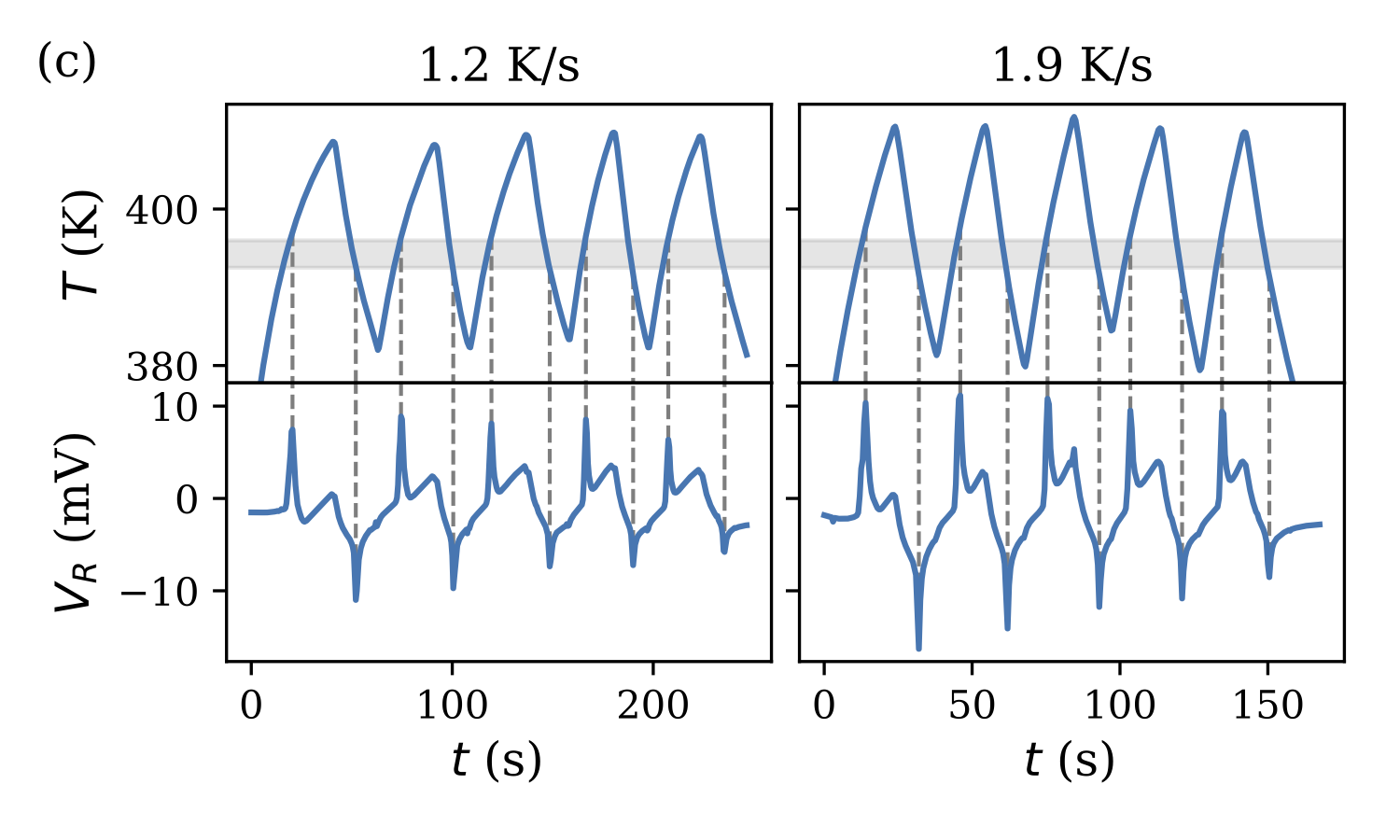}\\
\caption{Demonstration of voltage generation by the first-order phase transformation in (a) BTO, (b) Zr$_{0.006}$ and (c) Zr$_{0.01}$ at two different heating/cooling rates. In each subplot, upper panels are cyclic temperature in the specimen, while lower panels are the voltage across the resistor.
The shaded temperature ranges in upper panels are the same ranges as marked in FIG. \ref{fig:pyroelectric}.
}\label{fig:cycles}
\end{figure}

At $t=0$, we used a Tektronix PS280 DC power supply to charge the system and achieve the equilibrium voltage $\vb = 30$ V. At $t>0$, the power supply was detached and we monitor the voltage on the resistor while alternating the temperature of the system between 385 K and 405 K. Both the temperature and the voltage on resistor were recorded by an Agilent 34970A data acquisition/switch unit. Five consecutive transformation cycles of the system are shown in FIG. \ref{fig:cycles}. 

In FIG. \ref{fig:cycles}, we mark the same temperature range for phase transformation as in FIG. \ref{fig:pyroelectric}. 
Clearly the peaks of electricity signal, $V_{\rm R}$, exactly correspond to the temperature at which phase transformation occurs. This suggests that the electricity is generated solely and directly from the phase transformation of the material.  
Our results are different from the misleading electricity signal, for example in FIG. \ref{fig:schematics}b, whose peaks coincide with the alternation of the electric field $E$.

\subsection{Constitutive assumption}\label{sec:cons}

In order to solve \eqref{eq:ode-bar} numerically, 
we are going to propose a closed form constitutive relation $P(E, T)$.
Consider the case that the giant pyroelectric effect is caused by a first order phase transformation at the temperature $\theta(E)$. $\theta(E)$ is a function of electric field $E$ due to the Clausius-Clapeyron relation. 
It corresponds to the center of the decreasing slope in the $P-E$ curve under the applied field $E$.
Assuming a linear Clausius-Clapeyron relation 
\begin{equation}
\theta(E) =  \theta_0 + \xi E,    
\end{equation}
where $\xi$ is the \emph{Clausius-Clapeyron coefficient}.

First, we model the constitutive response for each of the single phases far away from phase transformation temperature by $P_{\rm f}(E)$ and $P_{\rm p}(E)$. Subscripts $_{\rm f}$ and $_{\rm p}$ denotes ferro- and para-electric phase respectively.
$P_{\rm p}(E)$ is always linear.
$P_{\rm f}(E)$ is linear under small field and approaches to the saturated value at high field.
Therefore we use the following hypothesis for
\begin{subequations}
\begin{align}
    P_{\rm f}(E) &= a_{\rm f} + b_{\rm f} \tanh(c_{\rm f} E), \label{eq:pff}\\
    P_{\rm p}(E) &= a_{\rm p} + \chi_{\rm p} E. \label{eq:pfp}
\end{align}
\end{subequations}
Here, $a$, $b$ $c$ and $\chi$ are constant coefficients. Naturally, define the \emph{jump} of polarization as
\begin{equation}
\jump{P}(E) = P_{\rm f}(E) - P_{\rm p}(E).    
\end{equation}

By the shape of constitutive response measured in FIG. \ref{fig:pyroelectric}, we model the function $P(E, T)$ as
\begin{equation}\label{eq:constitutive}
    P(E, T) = -\jump{P}(E) g(z(E, T)) + P_{\rm f}(E).
\end{equation}
where $\kappa$ is a material constant, $g(z)$ is the sigmoid function
\begin{equation}
g(z) = \frac{1}{1 + e^{-z}}.    
\end{equation}
and
\begin{equation}
z(E, T) = \frac{4 \kappa (T - \theta(E))}{\jump{P}(E)}.    
\end{equation}
Note that $g(E, T)$ and $z(E, T)$ are dimensionless.

The constitutive hypothesis \eqref{eq:constitutive} has the following properties:
\begin{itemize}
\item Under any $E$, $P \to P_{\rm f}(E)$ as $T \to 0$, and $P \to P_{\rm p}(E)$ as $T \to \infty$. Across the whole temperature range, $P$ monotonically decreases from $P_{\rm f}(E)$ to $P_{\rm p}(E)$.
\item Under any $E$, the steepest decreasing slope of $P-T$ occurs at $\theta(E)$ where the slope is $-\kappa$.
\item From $E = 0$, as $E$ increases, the $P-T$ curve shifts towards higher temperature, {\it i.e.} the Clausius-Clapeyron relation.
\end{itemize}

\begin{figure}
\centering
\includegraphics[width=3in]{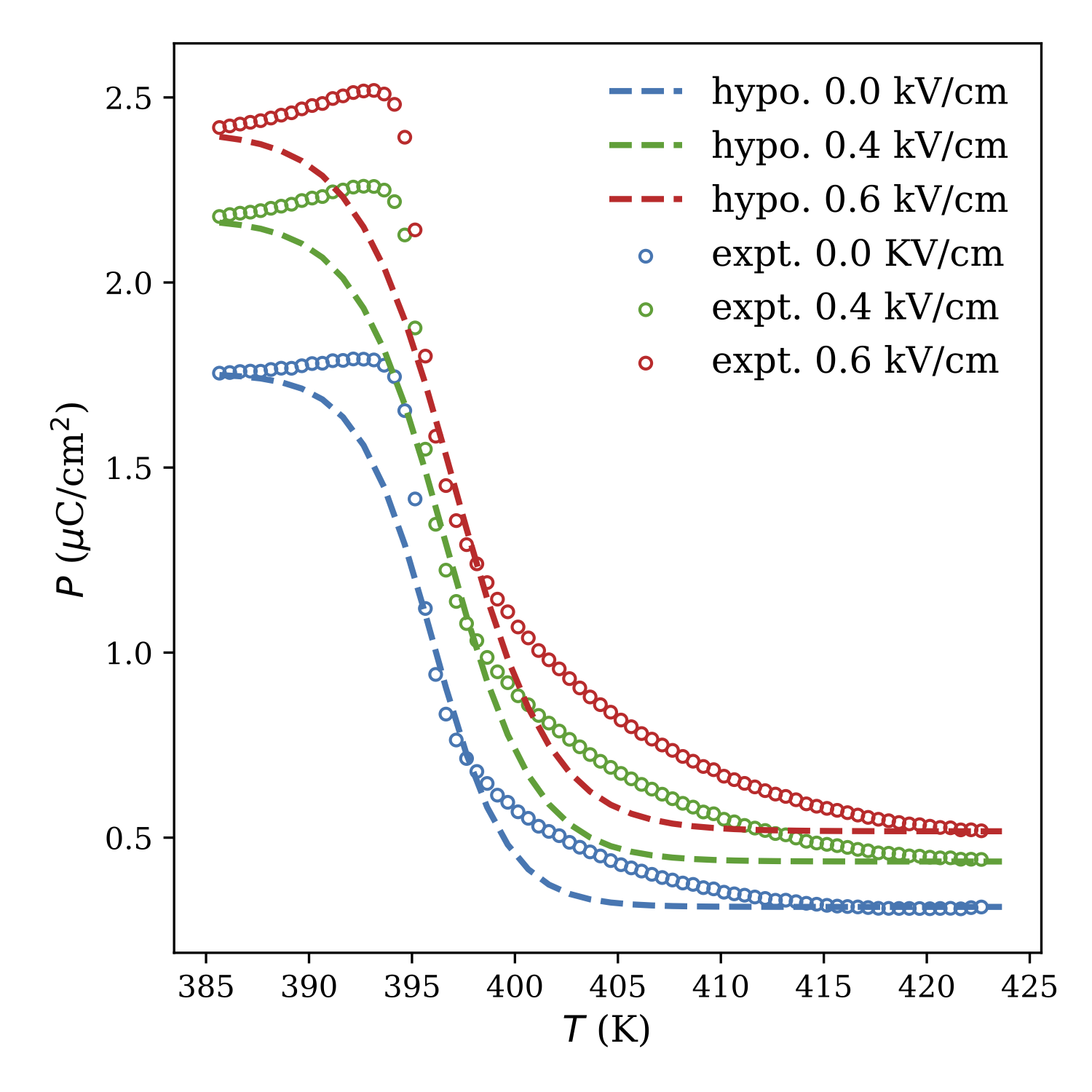}
\caption{Temperature dependency of polarization according to our constitutive hypothesis. Open circles are measured in the BTO specimen. Dashed lines are the hypothetical model as described in \eqref{eq:constitutive}.}\label{fig:PET}
\end{figure}

Using parameters listed in Table.\ref{tbl:params}, we evaluate this hypothetical model and compare with the experimental measurements for BTO in FIG.\ref{fig:PET}.
As shown in FIG.\ref{fig:PET}, our model captures the temperature dependency of the polarization within the phase transformation window reasonably well. 

\begin{table}
\centering
\begin{tabular}{ccl}
\hline
Variable & Value & Unit \\
\hline
 $\vb$ & 30 & V \\
 $\cref$ & 50 & $\mu$F \\
 $R$ & 100 & k$\Omega$ \\
 $A$ & 60 & mm$^2$ \\
 $d$ & 0.8 & mm \\
 & & \\
 $\theta_0$ & 396 & K \\
 $\kappa$ & 0.2 & $\mu$C/(cm$^2$ K) \\
 $\xi$ & 1.54 & K cm/kV \\
& & \\
 $a_{\rm f}$ & 1.7550 & $\mu$C/cm$^2$ \\
 $b_{\rm f}$ & 1.3042 & $\mu$C/cm$^2$ \\
 $c_{\rm f}$ & 0.8643 & cm/kV \\
 $a_{\rm p}$ & 0.3129 & $\mu$C/cm$^2$ \\
 $\chi_{\rm p}$ & 0.3146 & $\mu$C/(cm kV) \\
 & &\\
 $\tau$ & 10 & s \\
 \hline
\end{tabular}
\caption{Parameters used in simulation \label{tbl:params}}
\end{table}

Following the nondimensionalization introduced in Section \ref{sec:nondim},
we have
\begin{subequations}
\begin{align}
{\bar P}_{\rm f}(\bar E) &= \frac{1}{\subl P}\left(a_{\rm f} + b_{\rm f} \tanh\left(\frac{c_{\rm f}\vb\bar E}{d}\right)\right), \\
{\bar P}_{\rm p}(\bar E) &= \frac{1}{\subl P}\left(a_{\rm p} + \frac{\chi_{\rm p}\vb\bar E}{d}\right).
\end{align}
\end{subequations}
Rewrite the dimensionless function $z(E, T)$ as
\begin{equation}
z(\bar E, \bar T) =  \frac{4\kappa}{\subl P \jump{\bar P}(\bar E)}
\left(\frac{\subh T - \subl T}{2}\bar T + \frac{\xi\vb}{d}\left(1 - \bar E\right) \right).    
\end{equation}
The constitutive assumption \eqref{eq:constitutive} becomes
\begin{equation}
    \bar P(\bar E, \bar T) = -\jump{\bar P}(\bar E) g(z(\bar E, \bar T)) + {\bar P}_{\rm f}(\bar E).
\end{equation}

\subsection{Numerical simulation}

In real case, $\subl T$ and $\subh T$ are determined by the target working condition. 
To demonstrate a simulation of \eqref{eq:ode-bar}, we let
\begin{equation}
\begin{cases}
\subl T = \theta_0 + \dfrac{\xi V_{\rm b}}{d} - 8, \\
\subh T = \theta_0 + \dfrac{\xi V_{\rm b}}{d} + 2.
\end{cases}
\end{equation}
That is
\begin{equation}
\subl{\bar T} = -\frac{8}{5}, \quad \subh{\bar T} = \frac{2}{5}.    
\end{equation}
$\bar T(\bar t)$ is defined as
\begin{equation}
\bar T(\bar t) = - \cos\left(\pi\bar t\right) - \frac{3}{10}.    
\end{equation}

\begin{figure}[htpb]
\centering
\includegraphics[width=3in]{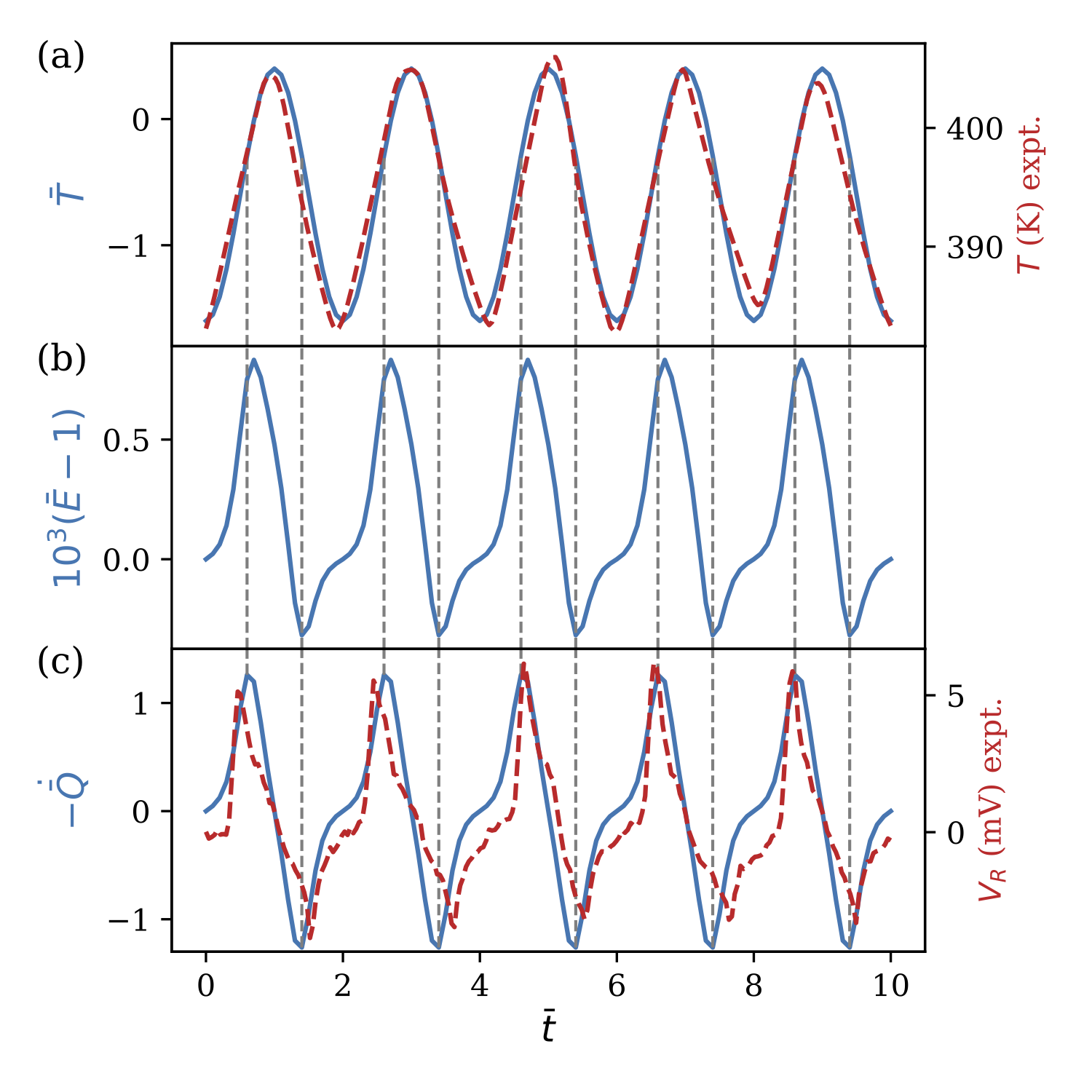}
\caption{
Simulated dimensionless (a) temperature, (b) field in pyroelectric material and (c) current flowing through the resistor in 5 consecutive cycles.
Blue solid lines are simulated results, red dashed lines are experimental measurements.
Dashed vertical lines indicate peaks in electricity signal, $-\dot{\bar Q}$.
\label{fig:cycles_sim}
}
\end{figure}

Using parameter values in TABLE~\ref{tbl:params}, we integrate \eqref{eq:ode-bar} from $\bar t = 0$ to $\bar t = 10$. Results are compared with experiments in FIG. \ref{fig:cycles_sim}. 
We show that our model captures the main features of behaviour of the battery detached energy conversion device. In particular, the peak signals  exactly appear at the phase transformation temperature, which implies the generated electricity is directly from heat.  
In simulated result FIG. \ref{fig:cycles_sim}c, the peak signals for heating and cooling are symmetric, while in the experiment FIG. \ref{fig:cycles}, the temperature corresponding to the heating peak is higher than that of the cooling peak. This is because we ignored the hysteresis of the first-order phase transformation in our constitutive hypothesis \eqref{eq:constitutive}.

\begin{figure}[ht]
\centering
\includegraphics[width=3in]{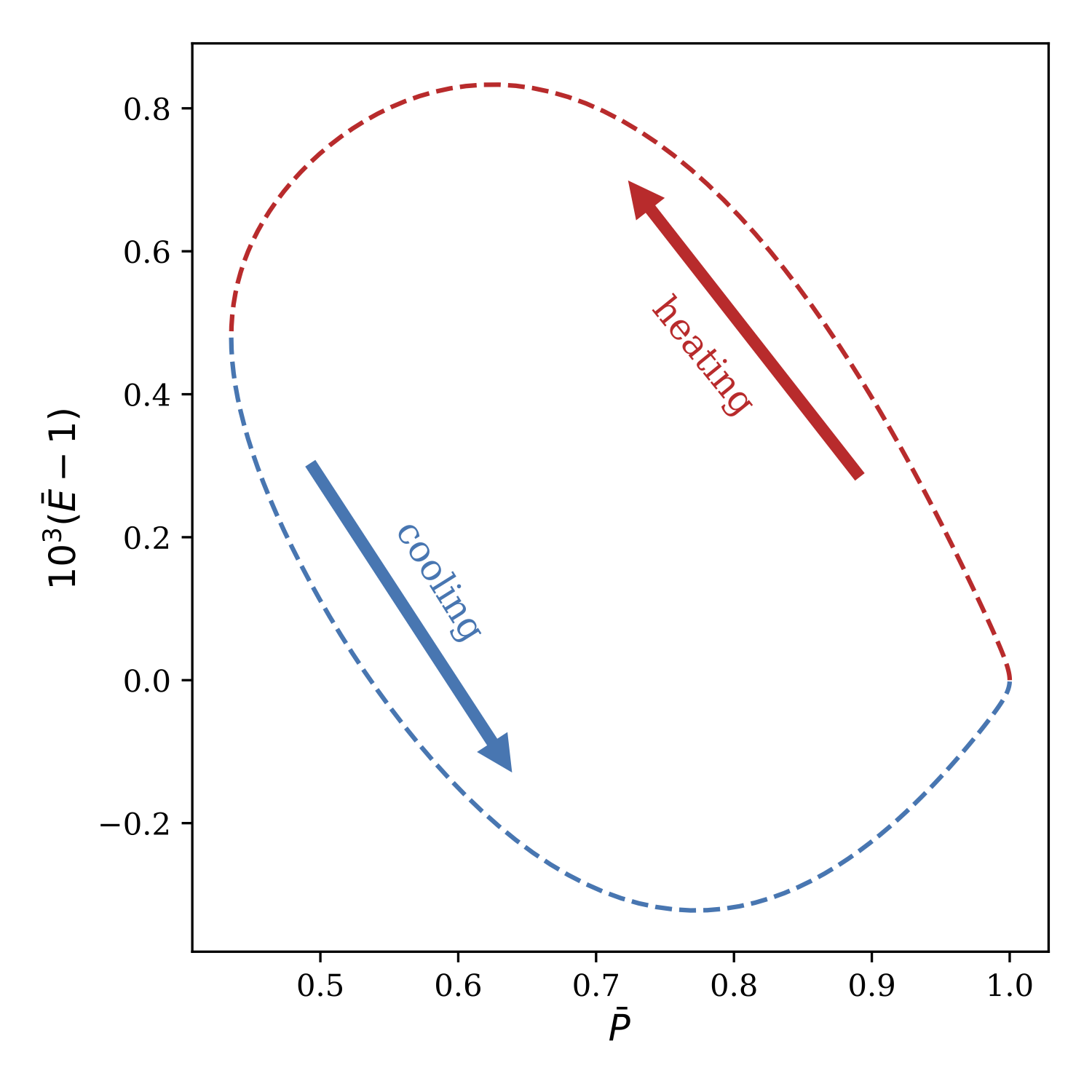}
\caption{
Simulated thermodynamic cycle in $P-E$ plane \label{fig:pe_sim}
}
\end{figure}

We draw the simulated thermodynamic cycle in $P-E$ plane in FIG. \ref{fig:pe_sim}.
The work output done by electrostatic work conjugate ($E$, $P$) is compared with the electricity output \eqref{eq:work}. Here, the electrostatic work is computed by
\begin{equation}\label{eq:work_es}
  \begin{aligned}
{\cal W}_{\rm{ES}} & = Ad \oint -E{\rm d}P = - Ad \int_{t_0}^{t_0 + 2\tau} E \dot P {\rm d}t \\
& = - \subl P\vb A\int_{\bar t_0}^{\bar t_0 + 2} \bar E \dot{\bar P} {\rm d}\bar t.
\end{aligned}  
\end{equation}
Choosing $\bar t_0 = 0$, we exactly have ${\cal W} = {\cal W}_{\rm{ES}}$.
In fact, one can prove that this equality holds analytically due to \eqref{eq:ode-bar}, \eqref{eq:work} and \eqref{eq:work_es}.
It means that all the work done by electrostatic force in the pyroelectric material is collected at the resistor.

\section{Further analysis}
\subsection{Approximated solution}
In this section we analyze the asymptotic limit of our pyroelectric energy conversion device. 
The dynamic equation \eqref{eq:ode-bar} can be converted to an ODE in terms of $\bar Q(\bar t)$ by changing variables. Then we are able to study the current term $\dot{\bar Q}(\bar t)$ directly.
Expand the Taylor series of $\bar Q(\bar E, \bar T)$ at $\bar Q(1, \bar T)$
\begin{equation}
\bar Q(\bar E, \bar T) = \bar Q(1, \bar T) + \bar Q_{,\bar E}(1, \bar T) (\bar E - 1) + h.o.t.    
\end{equation}
Here in $\bar Q_{,\bar E}$, the comma in the subscript denotes a partial derivative.
Ignoring higher order terms, \eqref{eq:ode-bar} can be written as a first order linear ODE for $\bar h = \bar Q_{,\bar E}(1, \bar T) (\bar E - 1)$:
\begin{equation}\label{eq:ode-h}
    \dot{\bar h} + \alpha(\bar t) \bar h + \beta(\bar t) = 0,
\end{equation}
where
\begin{equation}
\begin{cases}
\alpha(t) &= \left(1 + \dfrac{\vb\cref}{\subl P A \bar Q_{,\bar E}(1, \bar T)}\right)\dfrac{\tau}{R\cref}, \\
\beta(t) &=  \dot{\bar Q}(1, \bar T) + \dfrac{\tau}{R\cref}\bar Q(1, \bar T) 
- \dfrac{\tau}{R\cref}\left(\dfrac{\vb\epsilon_0}{\subl Pd} + 1\right) \\
&= {\bar P}_{,\bar T}(1, \bar T)\dot{\bar T} + \dfrac{\tau}{R\cref}\left(\bar P(1, \bar T) - 1\right).
\end{cases}
\end{equation}
The ODE \eqref{eq:ode-h} can be solved analytically by the method of integration factor. Here we ignore the detailed solving procedure and only focus on a special degeneracy of the equation.

In \eqref{eq:ode-h}, when $\alpha(t) \gg 1$ and $\alpha(t) \gg |\beta(t)|$, $\bar h$ asymptotically vanishes.
Then we can estimate $\bar Q(\bar E, \bar T)$ by $\bar Q(1, \bar T)$, which is rate independent.
By \eqref{eq:work}, the work output scales linearly with frequency: ${\cal W} \propto 1 / \tau$.

\begin{figure}[ht]
\centering
\includegraphics[width=3in]{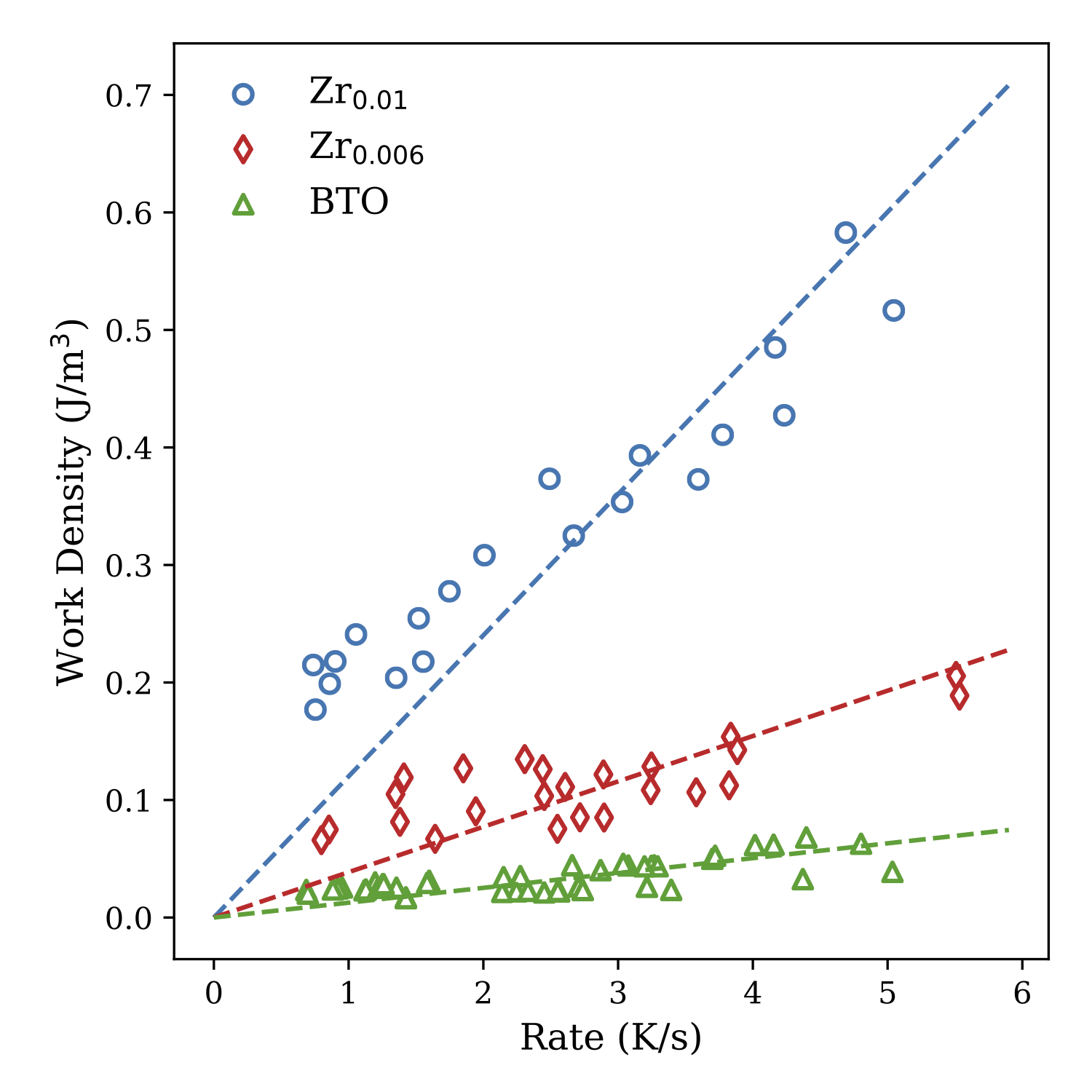}
\caption{Work output density per cycle at different heating/cooling rate. Each marker corresponds to one cycle from measurements. The first cycles in each measurement are excluded.}\label{fig:work}
\end{figure}

For our specimens, $\alpha$ is in the order of $10^6$ to $10^7$, $|\beta|$ is in the order of $1$ to $10$. 
The rate dependencies of work output per cycle are verified in our experiments for three testing samples shown in FIG. \ref{fig:work}. Compared to those state-of-art thin film pyroelectric devices assisted with alternated electric fields \cite{Olsen1985, Sebald2006, Lee2013, Pandya2018}, the maximum energy density purely converted from heat in bulk is in the same order of magnitude in Zr$_{0.01}$ specimen. 

\subsection{Threshold frequency}
In this section, we derive an estimated criterion under which the asymptotic conditions $\alpha(t) \gg 1$ and $\alpha(t) \gg |\beta(t)|$ hold.
First, we estimate $\bar Q_{,\bar E}(1, \bar T)$ as
\begin{equation}
\bar Q_{,\bar E}(1, \bar T) = \frac{\epsilon_0 \vb}{\subl P d} + \bar P_{,\bar E}(1, \bar T).    
\end{equation}
$\bar P_{,\bar E}$ comes from two factors: 1) single phase permittivity, 2) the coupling between the change in permittivity and Clausius-Clapeyron relation.
Denote ferroelectric susceptibility $\chi_{\rm f} = P_{\rm f}'(E)$. 
We have by \eqref{eq:pff}, $\chi_{\rm f} = b_{\rm f}c_{\rm f}\cosh^{-2}(c_{\rm f} E)$.
When $E$ is not too large, $\chi_{\rm f} \approx b_{\rm f}c_{\rm f}$.
As $E \to \infty$, $\chi_{\rm f} \to 0$.
Then we can estimate 
\begin{equation}
\bar Q_{,\bar E}(1, \bar T) \approx \frac{\vb}{\subl P d}\left(\epsilon_0 + \chi_{\rm f} + \kappa\xi \right).    
\end{equation}
That is
\begin{equation}\label{eq:alpha}
    \alpha \approx \frac{\tau}{R\cref}\left(1 + \frac{\cref d}{A (\epsilon_0 + \chi_{\rm f} + \kappa\xi)}\right),
\end{equation}
and 
\begin{equation}\label{eq:beta}
    \beta \approx \frac{\kappa (\subh T - \subl T)}{\subl P} + \frac{\tau}{R\cref}.
\end{equation}

From \eqref{eq:alpha}, to satisfy $\alpha \gg 1$, we need
\begin{equation}\label{eq:crit-1}
    \frac{\tau}{R\cref} \gg 1
\end{equation}
or
\begin{equation}\label{eq:crit-2}
    \frac{d\tau}{RA(\epsilon_0 + \chi_{\rm f} + \kappa\xi)} \gg 1.
\end{equation}
By \eqref{eq:beta}, if \eqref{eq:crit-1} holds, then $\alpha \gg \beta$ leads to
\begin{equation}
\frac{\cref d}{A (\epsilon_0 + \chi_{\rm f} + \kappa\xi)} \gg 1,
\end{equation}
which is not related to the rate parameter $\tau$.
If instead we have \eqref{eq:crit-2}, then
\begin{equation}
\alpha \gg \beta \implies \frac{\cref d}{A (\epsilon_0 + \chi_{\rm f} + \kappa\xi)} \gg \frac{\kappa (\subh T - \subl T)}{\subl P}.    
\end{equation}
Define the \emph{threshold frequency} to be\footnote[2]{The factor $1/10$ is chosen as a safety coefficient when considering the $\gg$ operator.}
\begin{equation}\label{eq:fstar}
    f^* = \frac{1}{10}\max\left\lbrace 
        \frac{1}{R\cref}, \frac{d}{RA(\epsilon_0 + \chi_{\rm f} + \kappa\xi)}
    \right\rbrace.
\end{equation}
When $1/\tau$ approaches or even exceeds $f^*$, the rate independent approximation $\bar Q(\bar E, \bar T) = \bar Q(1, \bar T)$ starts to fail.
$\cal W$ starts to scale only sub-linearly with $1/\tau$.

\begin{figure}[htpb]
\centering
\includegraphics[width=3in]{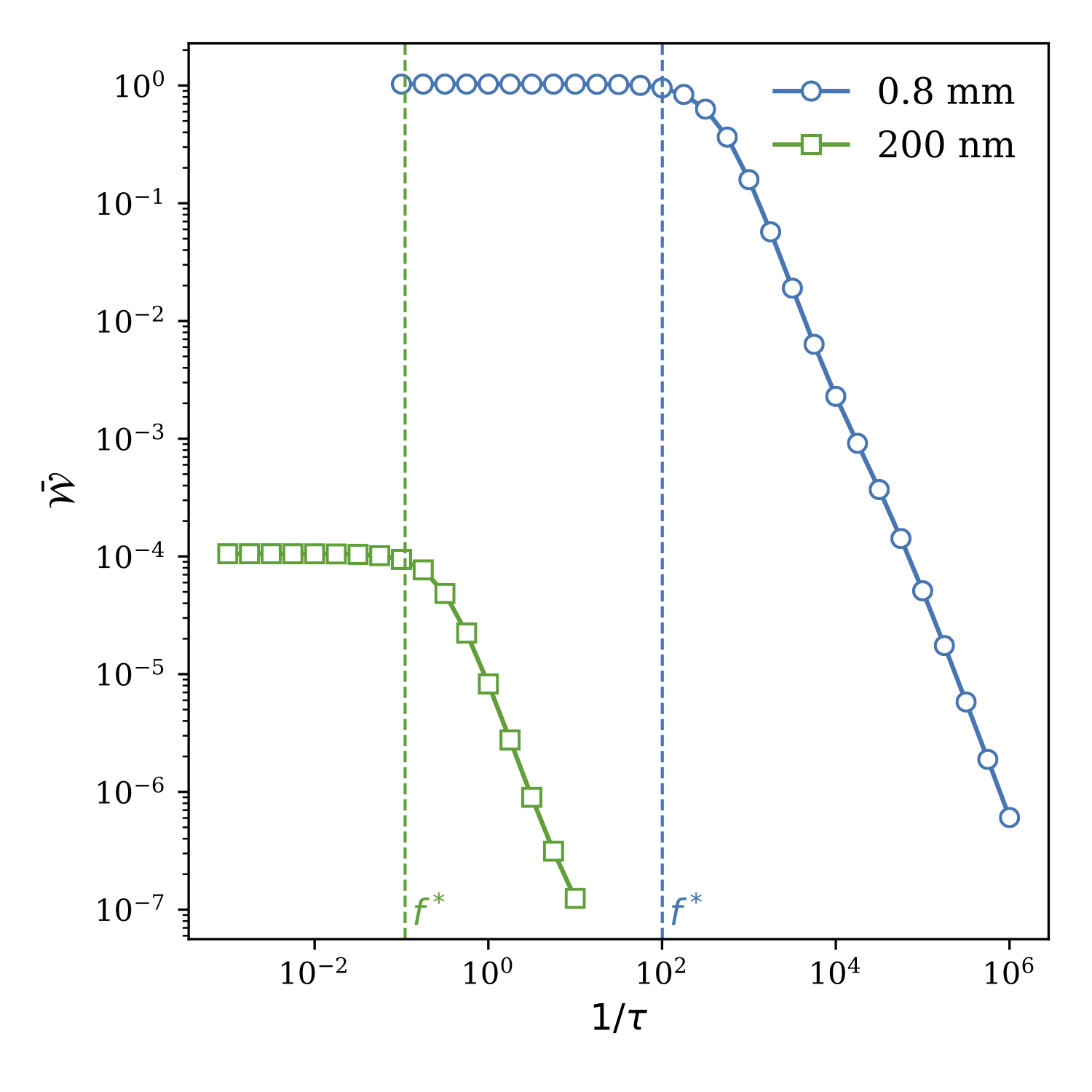}
\caption{
Simulated rate dependencies in bulk and thin film. The dimensionless work per cycle $\bar{\cal W}$ is evaluated for the 10$^{\rm th}$ cycle at each $\tau$. \label{fig:frequencies}
}
\end{figure}

From our constitutive model \eqref{eq:constitutive} and parameters in TABLE~\ref{tbl:params}, the threshold frequency $f^*$ = 100.6 Hz. So the rate independent approximation holds in almost all practical situations.\footnote[4]{Also, under a frequency as high as or even higher than 100 Hz, our pure direct-current model may need corrections.} This $f^*$ however will drastically decrease, if we use thin film geometry, as suggested by \eqref{eq:fstar}.
In FIG. \ref{fig:frequencies}, we show the simulated rate dependency on dimensionless work output $\bar{\cal W}$ for parameters in TABLE~\ref{tbl:params} and a thin film alternative with $d=200$ nm.
For the thin film, estimated $f^*$ = 0.11 Hz, the sub-linearity occurs at a much lower frequency compared to the bulk version. 
The dimensionless work output $\bar{\cal W}$ in the picture is defined as
\begin{equation}\label{eq:work-bar}
    \bar{\cal W} = \int_{\bar t_0}^{\bar t_0 + 2} \dot{\bar Q}^2 {\rm d}\bar t.
\end{equation}
By \eqref{eq:work} it is related to ${\cal W}$ via
\begin{equation}
    {\cal W} = \frac{\subl P^2 A^2 R}{\tau}\bar{\cal W}.
\end{equation}

\subsection{Figure of merit}
In this section, we are going to derive a simple estimate for the work output when $1/\tau$ is well below the threshold frequency,
\textit{i.e.} we can use the approximation $\bar Q(\bar E, \bar T) = \bar Q(1, \bar T)$.
Assuming the heating and cooling half cycles are symmetric, 
the work output of a full cycle is just twice that of the heating half cycle.
Considering the dimensionless work \eqref{eq:work-bar},
in our approximation, it reads
\begin{equation}
\begin{aligned}
\bar{\cal W} & = 2\int_{\bar t_0}^{\bar t_0 + 1} \dot{\bar Q}(1, \bar T)^2 {\rm d}\bar t = 2\int_{\bar t_0}^{\bar t_0 + 1} \dot{\bar P}(1, \bar T)^2 {\rm d}\bar t \\
& = 2\int_{\bar t_0}^{\bar t_0 + 1} \dot{\bar P}(1, \bar T) {\bar P}_{,\bar T}(1, \bar T) \dot{\bar T} {\rm d}\bar t \\
& = 2\int_{\subl{\bar T}}^{\subh{\bar T}} \dot{\bar P}(1, \bar T) {\bar P}_{,\bar T}(1, \bar T) {\rm d}\bar T.
\end{aligned}    
\end{equation}
Since ${\bar P}_{,\bar T}(1, \bar T)$ is always negative in the temperature range $[\subl{\bar T}, \subh{\bar T}]$. By mean value theorem, there exists a value $\subl{\bar T} \leqslant \gamma \leqslant \subh{\bar T}$ such that
\begin{equation}\label{eq:work-bar-2}
    \begin{aligned}
    \bar{\cal W} & = 2\dot{\bar P}(1, \gamma) \int_{\subl{\bar T}}^{\subh{\bar T}} {\bar P}_{,\bar T}(1, \bar T) {\rm d}\bar T \\
    & = 2{\bar P}_{,\bar T}(1, \gamma)\dot{\bar T}(\gamma)
    ({\bar P}(1, \subh T) - {\bar P}(1, \subl T)).
    \end{aligned}
\end{equation}

Up to now, the derivation is exact. Next we estimate each term in \eqref{eq:work-bar-2}:
\begin{equation}
\begin{cases}
&{\bar P}_{,\bar T}(1, \gamma) \approx -\dfrac{\kappa(\subh T - \subl T)}{2\subl P}, \\
&\dot{\bar T}(\gamma) \approx \dfrac{2}{\subh T - \subl T}, \\
&{\bar P}(1, \subh T) - {\bar P}(1, \subl T) \approx
-\jump{\bar{P}}(1).
\end{cases}
\end{equation}
Substituting back into \eqref{eq:work} gives
\begin{equation}
    {\cal W} \approx \frac{2\kappa\jump{P}A^2R}{\tau},
\end{equation}
where $\jump{P}$ is the short hand of the functional evaluation $\jump{P}(\vb/d)$, which can be loosely interpreted as ``the jump of polarization across phase transformation''.
The work density is then
\begin{equation}
    w \approx \frac{2\kappa\jump{P}AR}{d\tau}.
\end{equation}
Recall, as discussed in Section \ref{sec:nondim}, that when the heat flow is pre-determined and $[\subl{T}, \subh{T}]$ covers the whole phase transformation region, $\tau \propto \ell$, where $\ell$ is the latent heat of the phase transformation.
We propose the figure of merit (FoM)
\begin{equation}\label{eq:fom}
    \zeta = \frac{\kappa\jump{P}}{\ell}\frac{AR}{d}
    := \zeta_1\zeta_2.
\end{equation}
Here, $\zeta_1$ is determined by material properties, while $\zeta_2$ depends on device parameters. For our specimens, the material FoM $\zeta_1$ is 0.35, 3.5 and 4.0 $\mu$C$^2$/(J cm K) for BTO, Zr$_{0.006}$ and Zr$_{0.01}$ respectively, which agree well with the performance measured at various rates in FIG.~\ref{fig:work}. Consider the efficiency estimated as the ratio of the work density and the absorbed heat, an equivalent material FoM can be expressed as $\frac{\kappa \jump{P}}{\ell^2}$. 
Conventionally, people usually use the pyroelectric coefficient $\kappa = P_{,T}$ as a critical performance measure of material exploited for pyroelectric energy conversion. According to \eqref{eq:fom}, both pyroelectric property and the thermal property of the material should be considered for material development. However, this does not mean that the second order phase transformation of much smaller latent heat is superior to the first order phase transformation. Suppose two materials having the same efficiency, the one with larger latent heat will release more energy during the ferro- to para-electric phase transformation.
When an energy system only concerns the maximum energy output, the material FoM $\zeta_1$ is the major factor, e.g. that Zr$_{0.006}$ and Zr$_{0.01}$ have much better performance than BTO in FIG.~\ref{fig:work}. This result inspires further material research and device engineering to achieve better energy conversion performance.


\begin{acknowledgments}
C. Z. and X. C. thank the financial support of the HK Research Grants Council under Grant 26200316 and 16207017. X. C. also thanks the Isaac Newton Institute for Mathematical Sciences for support and hospitality during the programme ``The mathematical design of new materials" when work on this paper was undertaken. This work was supported by EPSRC grant number EP/R014604/1. E.Q. and M.W. acknowledge funding by the DFG through a Reinhart Koselleck project. 
\end{acknowledgments}

\bibliographystyle{apsrmp4-1}
\bibliography{energy}

\end{document}